\documentclass[12pt]{article}
\pdfoutput=1

\usepackage[toc,page]{appendix}
\usepackage{slashed}
\usepackage[sumlimits,intlimits,namelimits]{amsmath} 
\usepackage[english]{babel}
\usepackage{graphicx}
\usepackage{hyperref}
\usepackage{cite}

\newcommand{\ba}{\begin{eqnarray}}
\newcommand{\ea}{\end{eqnarray}}

\newcommand{\ice}[1]{\relax}

\usepackage{color}
\usepackage{amsmath}
\usepackage{booktabs} 
\usepackage{subfigure}

\begin{document}

\begin{titlepage}
  
\begin{flushright}
SI-HEP-2023-07\\
SFB-257-P3H-23-24\\
PSI-PR-23-10
\end{flushright}
\vspace{0.12cm}
\begin{center}
	  { \Large\bf The heavy quark expansion for lifetimes: \\ [2mm]
	  	          Towards the QCD corrections to power suppressed terms} 
\end{center}
\vspace{0.35cm}
\begin{center}
  {\sc Thomas Mannel${}^1$}, {\sc Daniel Moreno${}^2$},
  and {\sc Alexei A. Pivovarov${}^1$} \\[0.2cm]
  {\sf ${}^1$Center for Particle Physics Siegen,
    Theoretische Physik 1, Universit\"at Siegen\\ 57068 Siegen, Germany} \\[1mm] 
  {\sf ${}^2$Paul Scherrer Institut, CH-5232 Villigen PSI, Switzerland}
\end{center}

\vspace{0.48cm}
\begin{abstract}\noindent
We consider the Heavy Quark Expansion (HQE) for the nonleptonic decay rates
of heavy hadrons, and compute the NLO QCD corrections to power terms up to order $1/m_Q^2$.
We neglect the masses of the final-state quarks, so the application of our result is mainly for charmed hadrons.
Our result can be applied also to bottomed hadrons as they constitute the main effect to this order up to corrections
of $\mathcal{O}(m_c/m_b)$ and contributions due to penguin operators.
We discuss the impact of our result for the lifetimes of heavy hadrons.
\end{abstract}
\end{titlepage}

\section{Introduction} 
\label{sec:Intro}
With the development of the heavy quark expansion (HQE)~\cite{Shifman:1987rj,Eichten:1989zv,Isgur:1989vq,Grinstein:1990mj},
the theoretical description of inclusive decay rates of heavy hadrons (i.e. of hadrons containing a single heavy quark $Q$)
has been advanced significantly. The HQE allows us to describe their decay rates $\Gamma$ and
spectra as a systematic expansion of the form~\cite{Chay:1990da,Bigi:1992su,Bigi:1993fe,Blok:1993va,Manohar:1993qn}

\begin{equation}
 \Gamma = \sum_{n=0}^{\infty}\Gamma_n \bigg(\frac{1}{m_Q}\bigg)^n\,,
\end{equation}
where the $\Gamma_n \propto \Lambda_{\rm QCD}^n$ involve non-perturbative parameters, the so called HQE parameters, with
coefficients that can be computed perturbatively as a power series in $\alpha_s (m_Q)$.

Over the last decades, this method has been continuously improved and refined, in particular by
computing higher-order corrections in $\Lambda_{\rm QCD} / m_Q$  as well as higher-orders in 
$\alpha_s (m_Q)$. For inclusive semileptonic decays and motivated by the possibility to determine 
$V_{cb}$ with a high precision, the HQE has been investigated very intensively, while for inclusive nonleptonic
rates the HQE has been pushed to a similar level. 

The most inclusive quantities are the lifetimes of heavy hadrons, which can be computed in the HQE. Its main
prediction is that the leading contribution to the heavy hadron lifetime is described by the decay rate
of the corresponding free heavy quark. To this end, the HQE thus predicts that all heavy-hadron lifetimes are equal up to
corrections of order $(\Lambda_{\rm QCD} / m_Q)^2$, since the term linear in the expansion
parameter is absent due to heavy quark symmetries. In fact this was an embarrassment in the early
days of the HQE, since at that time only measurements of lifetimes of charmed hadrons were available. 
The current numbers are~\cite{HFLAV:2022pwe}

\begin{equation}
	\label{eq:charm-lifetimes-exper}
	\frac{\tau(D^\pm)}{\tau(D^0)}\bigg|^{{\scriptsize\mbox{exp}}} = 2.563 \pm 0.017\,,
	\;\;\;\; 
	\frac{\tau(D_s)}{\tau(D^0)}\bigg|^{{\scriptsize\mbox{exp}}} = 1.219 \pm 0.017\,,
		\;\;\;\; 
	\frac{\tau(D^\pm)}{\tau(\Lambda_c)}\bigg|^{{\scriptsize\mbox{exp}}} = 5.123 \pm 0.014\,,
\end{equation}
which are in contrast to the expectation of a few percent. This clearly shows that this simple 
picture is too naive in the case of charm, leaving us with some doubt on the applicability of the
HQE for the charm quark~\cite{Mannel:2021uoz}. Within the HQE, the large lifetime differences are tracked by matrix elements
of four quark operators which have Wilson coefficients
that are enhanced by a $16\pi^2$ phase-space factor and scale as $16\pi^2 (\Lambda_{\rm QCD}/m_c)^3$ relative to the leading term~\cite{Neubert:1996we}.
In the case of charm, these terms can become comparable to the leading term.
The successful applications of the HQE to charm are all
related to observables where the matrix elements of these four quark operators are suppressed for some physical reason.
The HQE for charmed hadrons have been extensively used to explore its applicability, e.g in~\cite{King:2021xqp,King:2021jsq,Gratrex:2022xpm,Cheng:2023jpz,Dulibic:2023jeu}.

In contrast, for the bottom quark this picture seems to be more realistic, since one finds for the bottom
hadrons~\cite{HFLAV:2022pwe}
\begin{equation}
	\frac{\tau(B_s)}{\tau(B_d)}\bigg|^{{\mbox{\scriptsize exp}}} = 0.998 \pm 0.004\,,
	\;\;\;\; 
	\frac{\tau(B^+)}{\tau(B_d)}\bigg|^{{\scriptsize\mbox{exp}}} = 1.076 \pm 0.004\,
		\;\;\;\; 
	\frac{\tau(\Lambda_b)}{\tau(B^+)}\bigg|^{{\scriptsize\mbox{exp}}} = 0.969 \pm 0.006\,
\end{equation}
which is a clear motivation for considering also higher order corrections to the HQE
of lifetimes, whose current status has been
presented in~\cite{Lenz:2014jha,Cheng:2018rkz,Lenz:2022rbq,Gratrex:2023pfn,Piscopo:2023jnu}.

As for the current knowledge of the perturbative QCD corrections to the coefficients of the HQE of the rate,
the situation is the following:

\begin{itemize}
 \item {\bf Semileptonic decays}: The leading power coefficient is known at N$^3$LO~\cite{Fael:2020tow,Czakon:2021ybq}. The coefficients of the first power correction are known at NLO~\cite{Alberti:2013kxa,Mannel:2014xza,Mannel:2015jka}.
 From the second power correction onwards four-quark operators start to appear.
 For the second power correction the coefficients of the two-quark and four-quark operators are known at
 NLO~\cite{Mannel:2019qel,Mannel:2021zzr,Moreno:2022goo,Lenz:2013aua}. Finally, the coefficients of the third and fourth power corrections
 are known at LO~\cite{Dassinger:2006md,Mannel:2010wj} for the two-quark operators.

 \item {\bf Nonleptonic decays}: The leading power coefficient is known at
 NLO~\cite{Altarelli:1980fi,Buchalla:1992gc,Bagan:1994zd,Krinner:2013cja} and at NNLO in
 the massless case for the color-singlet $\Delta B=1$ operator~\cite{Czarnecki:2005vr}. The coefficients of the first power correction are known at
 LO~\cite{Bigi:1992su,Blok:1992hw,Blok:1992he}. The coefficients of the second
 power correction are known at LO for the two-quark operators~\cite{Lenz:2020oce,Mannel:2020fts,Moreno:2020rmk}
 and at NLO for the four-quark operators~\cite{Beneke:2002rj,Mescia:2002}. Finally, the coefficients of the third power
 correction are known at LO for the four-quark operators~\cite{Gabbiani:2004tp}.
\end{itemize}

In the present paper we extend the existing calculations for nonleptonic widths by computing $\alpha_s$ corrections
to power suppressed terms at next-to-leading power. We present an analytical result for the nonleptonic width
at order $\alpha_s(m_Q) (\Lambda_{\rm QCD}/m_Q)^2$ for the case of vanishing final state quark masses.

The main application of our result is $D$ hadron decays as it corresponds
to the Cabibbo-Kobayashi-Maskawa (CKM) favoured decay channel $c\rightarrow s\bar d u$.
To some extent, our results can be applied to $B$ hadron decays. To order $\alpha_s(m_Q) (\Lambda_{\rm QCD}/m_Q)^2$ they
constitute the main effect in the CKM favoured
decay channel $b\rightarrow c\bar u d$ up to corrections of $\mathcal{O}(m_c/m_b)$. The same is true
for the CKM favoured decay channel $b\rightarrow c\bar c s$ up to corrections of $\mathcal{O}(m_c/m_b)$ and
up to the effect of penguin operators, which is not considered in this paper.

The paper is organized as follows.
In section~\ref{sec:LeffEW} we discuss the effective electroweak Lagrangian and the choice of the
renormalization scheme. In section~\ref{sec:hqerate} we set the definitions for the HQE.
In section~\ref{sec:outline} we describe our method for the computation. Finally, we collect the results
and discuss their impact in section~\ref{sec:res}.

\section{The effective electroweak Lagrangian}
\label{sec:LeffEW}

In this section we discuss the effective Lagrangian describing nonleptonic transitions and provide the main
definitions needed for this paper. At low momentum transfer compared to the $W$-boson mass $M_W$, the nonleptonic heavy quark decay
$Q\rightarrow q_1 \bar q_2 q_3$ can be described by an effective Fermi Lagrangian
\begin{equation}
 \mathcal{L}_{{\scriptsize\mbox{eff}}} = - 2\sqrt{2} G_F V_{q_2 q_3} V_{q_1 Q}^* (C_1 \mathcal{O}_1 + C_2 \mathcal{O}_2)
 + \mbox{h.c}\,,
 \label{eq:FermiLagr}
\end{equation}
where $G_F$ is the Fermi constant, $V_{qq'}$ are the corresponding matrix elements of the CKM matrix and $C_{1,2}$ are matching
coefficients. We start from the standard operator basis $\mathcal{O}_{1,2}$ with color singlet and color rearranged operators~\cite{Buras:1989xd}
\begin{eqnarray}
\label{basis}
 \mathcal{O}_1 &=& (\bar Q^i \Gamma_\mu q_1^j)(\bar q_2^j \Gamma^\mu q_3^i)\,,
 \\
 \mathcal{O}_2 &=& (\bar Q^i \Gamma_\mu q_1^i)(\bar q_2^j \Gamma^\mu q_3^j)\,,
\end{eqnarray}
where $\Gamma_\mu=\gamma_\mu (1-\gamma_5)/2=\gamma_\mu P_L$, $(i,j)$ are color indices, and $q_{1,2,3}$ are the final state quarks
which we take to be massless in the following. We assume for simplicity that the three final-state quarks have different
flavors, so we do not need to consider QCD penguin operators.

However, for the calculation we address in this paper, it is convenient to chose a different operator basis for our effective Lagrangian
in Eq.~(\ref{eq:FermiLagr})
\begin{eqnarray}
 {\cal L}_{\rm eff} = -2\sqrt{2}G_F V_{q_2 q_3} V_{q_1 Q}^* ( C_{+} \mathcal{O}_{+} + C_{-} \mathcal{O}_{-})
 + \mbox{h.c}\,,
 \label{LeffEWpm}
\end{eqnarray}
with $\mathcal{O}_\pm = (\mathcal{O}_2 \pm \mathcal{O}_1)/2$ and $C_\pm = C_2 \pm C_1$.
The advantage is that this basis is diagonal under renormalization. In the $\overline{\mbox{MS}}$ renormalization scheme
\begin{eqnarray}
 C_{\pm\,,B} = Z_{\pm} C_\pm \,,\quad\quad
 Z_{\pm} = 1 + \frac{1}{2} \gamma_{\pm} \frac{\alpha_s(\mu)}{4\pi}\frac{1}{\epsilon}\,,\quad\quad
 \gamma_\pm = - 6\bigg(\frac{1}{N_c} \mp 1\bigg)\,.
\end{eqnarray}
where the subindex $B$ stands for bare quantities and those without subscript stand for renormalized ones,
$N_c=3$ is the number of colors and $\gamma_{\pm}$ is the LO anomalous dimension of the operators $\mathcal{O}_\pm$.

An important technical issue here is to retain the same scheme for the calculation of correlators and for the
calculation of the short distance Wilson coefficients $C_{\pm}$ appearing in the effective Lagrangian.
The point is that the renormalization of the operators $\mathcal{O}_{\pm}$ is additionally
complicated by the fact that they involve left-handed fields and require the special treatment of $\gamma_5$ in dimensional
regularization. There are several possibilities like dimensional reduction~\cite{Altarelli:1980fi}, the
't Hooft-Veltman scheme~\cite{Buchalla:1992gc}, or Naive Dimensional Regularization (NDR) with
anticommuting $\gamma_5$~\cite{Bagan:1994zd}.

We decide to closely follow the approach used by \cite{Bagan:1994zd} and chose to work in NDR within the scheme of evanescent
operators that preserves Fierz symmetry~\cite{Buras:1989xd,Dugan:1990df,Herrlich:1994kh,Jamin:1994sv,Grozin:2018wtg,Grozin:2017uto,Grozin:2016uqy}. Such evanescent operators are defined in ref.~\cite{Buras:1989xd}, where the
two-loop anomalous dimension required for the running of $C_\pm$ at NLO is also computed.
This definition respects the Fierz transformation which in general is valid only in four-dimensional space-time.
This choice is very handy as it allows, by using an appropriate Fierz transformation, to avoid closed fermionic loops, which are known to lead to algebraic inconsistencies when using anticommuting $\gamma_5$ in $D$ dimensions.

The freedom in the choice of evanescent operators is connected to the freedom in the choice of the renormalization scheme.
Such a freedom is represented by the shift
\begin{eqnarray}
A\gamma_\pm = A\bigg(\pm 1+\frac{1}{N_c}\bigg)\,,
\end{eqnarray}
proportional to the LO anomalous dimensions $\gamma_\pm$ of the operators $\mathcal{O}_\pm$.

In the following we give the definition for the coefficients $C_\pm$ in NDR within the scheme of evanescent
operators that preserves Fierz symmetry. The Wilson coefficients $C_{\pm}$ with NLO precision
(including also the renormalization group improvement at NLO) are given by~\cite{Altarelli:1980fi,Buras:1989xd,Bagan:1994zd}
\begin{eqnarray}
	C_{\pm}(\mu) = L_{\pm}(\mu)\bigg[1 + \frac{\alpha_s(M_W) - \alpha_s(\mu)}{4\pi}R_{\pm} + \frac{\alpha_s(\mu)}{4\pi}B_{\pm}\bigg]\,,
	\label{CpmNLO}
\end{eqnarray}
which have been calculated at the scale $\mu = M_W$ and then evolved down to scales $\mu \ll M_W$ by solving the corresponding renormalization group equations. The equation above splits the coefficients into a scheme-independent part proportional to $R_{\pm}$
and a scheme dependent part proportional to $B_{\pm}$, with~\cite{Altarelli:1980fi,Buras:1989xd,Buchalla:1992gc,Bagan:1994zd}
\begin{eqnarray}
	R_{+} = \frac{10863 - 1278n_f + 80n_f^2}{6(33-2n_f)^2}\,,\quad
	R_{-} = - \frac{15021 - 1530n_f + 80n_f^2}{3(33-2n_f)^2}\,, \quad
	B_{\pm} = \frac{1}{12} B \gamma_\pm \,,
	\label{RpmBpm}
\end{eqnarray}
where $n_f$ is the number of light flavours and $B=11$ in NDR with anticommuting $\gamma_5$~\cite{Buras:1989xd}. The last equation is implied by Fierz symmetry.
The matching coefficients $B_{\pm}$ ensure that, up to terms of order $\alpha_s^2(M_W)$,
matrix elements of the effective Lagrangian calculated at the scale $\mu=M_W$ are equal to the
corresponding matrix elements calculated with the full standard model Lagrangian.
Eventually, the scheme-dependence absorbed in $B_{\pm}$ has to cancel
against the scheme-dependence of matrix elements of the corresponding operators.

Finally,
\begin{equation}
	L_{\pm}(\mu) = \bigg(\frac{\alpha_s(M_W)}{\alpha_s(\mu)}\bigg)^{\frac{\gamma_\pm}{2\beta_0}}\,,
	\label{lpm}
\end{equation}
is the solution of the RGE for $C_{\pm}$ to leading logarithmic accuracy, with
$\beta_0 = \frac{11}{3}N_c - \frac{2}{3} n_f$.

\section{HQE for nonleptonic decays of heavy flavors}
\label{sec:hqerate}

This section briefly describes the theoretical framework used for the calculation of inclusive nonleptonic decays of heavy hadrons
within the HQE and provides the main definitions. We follow the approach introduced in~\cite{Mannel:2014xza,Mannel:2015jka,Mannel:2021zzr}.

By using the optical theorem one obtains the inclusive decay rate $\Gamma$
from taking an absorptive part of the forward matrix element of the leading order transition 
operator ${\cal T}$
\begin{equation}\label{eq:trans_operator}
	{\cal T} = i\, \int d^4 x\,
	T\{ {\cal L}_{\rm eff} (x)  {\cal L}_{\rm eff} (0) \} \, ,
	\quad\quad  \Gamma (H_Q \to X)
	= \frac{1}{M_{H_Q}} \text{Im } \langle H_Q|{\cal T} |H_Q\rangle \, ,
\end{equation} 
where $M_{H_Q}$ is the heavy hadron mass and $|H_Q\rangle$ its quantum state. Since the heavy quark mass
$m_Q$ is a large scale compared to the QCD hadronization
scale  $\Lambda_{\rm QCD}$ ($m_Q\gg\Lambda_{\rm QCD}$), the
forward matrix element contains perturbatively calculable 
contributions. These can be separated from the non-perturbative
pieces using the method of effective field theory.  
For a heavy hadron with momentum $p_{H_Q}$ and mass $M_{H_Q}$, a large part
of the heavy-quark momentum $p_Q$ originates from a pure kinematical
contribution due to its large mass. We split the heavy-quark momentum
according to $p_Q=m_Q v+\Delta$ with $v=p_{H_Q}/M_{H_Q}$ being the velocity of
the heavy hadron. 
The residual momentum $\Delta \sim {\cal O} (\Lambda_{\rm QCD})$
describes 
the soft-scale fluctuations 
of the heavy quark field near its mass shell.

This decomposition of the quark momentum is implemented 
by re-defining the heavy quark field according to 
\begin{equation}\label{eq:heavy-quark-me-phase}
	Q(x) = e^{-i m_Q v\cdot x}Q_v (x)\, ,
\end{equation}
so that $i \partial Q_v (x) \sim \Delta$. 

We set up the HQE as an expansion in $\Lambda_{\rm QCD}/m_Q$ by matching the
transition operator ${\cal T}$ in QCD to an expansion in inverse powers of the heavy quark mass, 
using operators defined in Heavy Quark Effective Theory (HQET)~\cite{Mannel:1991mc,Manohar:1997qy,Georgi:1990um,Neubert:1993mb}.

Generally, the HQE for heavy hadron weak decays takes the form
\begin{equation}
	\label{hqewidth}
	\Gamma(H_Q\rightarrow X) = \Gamma^0 |V_{q_2 q_3}|^2 |V_{q_1 Q}|^2
	\bigg(  C_0
	- C_{\mu_\pi}\frac{\mu_\pi^2}{2m_Q^2}
	+ C_{\mu_G}\frac{\mu_G^2}{2m_Q^2} + \cdots
	\bigg)\,,
\end{equation}
where 
$\Gamma^0= G_F^2 m_Q^5/(192 \pi^3)$ and the ellipses denote terms of order $1/m_Q^n$, $n \ge 3$.
The coefficients $C_0$, $C_{\mu_\pi}$ and $C_{\mu_G}$ can be computed as a power series in $\alpha_s(\mu)$ and depend,
in case of neglecting the final-state quark masses, on logarithms of $\mu/m_Q$, where $\mu$ is the matching scale.
Therefore, for $\mu=m_Q$ the coefficients are pure numbers. The parameters $\mu_\pi^2$, $\mu_G^2$ are forward matrix elements
of local HQET operators called HQE parameters.

The previous expression emerges from the direct matching of
the QCD expression for the transition operator to HQET   
\begin{eqnarray}
	\label{eq:HQE-1}
	\mbox{Im } \mathcal{T} = \Gamma^0 |V_{q_2 q_3}|^2 |V_{q_1 Q}|^2
	\bigg( C_0 \mathcal{O}_0 
	+ C_v \frac{\mathcal{O}_v}{m_Q}
	+ C_\pi \frac{\mathcal{O}_\pi}{2m_Q^2}
	+ C_G \frac{\mathcal{O}_G}{2m_Q^2} + \cdots
	\bigg)\,,
\end{eqnarray}
where again the coefficients $C_0$, $C_v$, $C_\pi$ and $C_G$ can be computed as a power series in $\alpha_s(\mu)$.
The local operators~${\cal O}_i$ in the equation above are
ordered by their mass dimensionality and are given by\footnote{In general, there is an additional operator 
	${\cal O}_{\rm{I}} = \bar{h}_v (v\cdot\pi)^2 h_v $ in the complete basis at dimension
	five. However it will be of higher order in the HQE
	after using equations of motion of HQET.}
\begin{eqnarray}
	\mathcal{O}_0 &=& \bar h_v h_v \qquad\qquad\qquad\qquad\qquad\quad\;\;\, \mbox{(mass dimension three)}\,,
	\label{O0}
	\\
	\mathcal{O}_v &=& \bar h_v v\cdot \pi h_v \quad\qquad\qquad\qquad\qquad\;\;\; \mbox{(mass dimension four)}\,,
	\label{Ov}
	\\
	\mathcal{O}_\pi &=& \bar h_v \pi_\perp^2 h_v
	\qquad\qquad\qquad\qquad\quad\quad\;\;
	\mbox{(mass dimension five)}\,,
	\label{mupi} \\
	\mathcal{O}_G &=& \frac{1}{2}\bar h_v [\gamma^\mu, \gamma^\nu]
	\pi_{\perp\,\mu}\pi_{\perp\,\nu}  h_v
	\quad\quad\qquad\;\; \mbox{(mass dimension five)}\,,
	\label{OG}
\end{eqnarray}
where $\pi_\mu = i D_\mu = i\partial_\mu +g_s A_\mu^a T^a$ is the
covariant derivative of QCD and 
$\pi^\mu =v^\mu (v\pi)+\pi^\mu_\perp$.

Note that the field $h_v$ denotes the static quark field moving with the velocity $v$ as
defined in HQET. Furthermore, it is convenient to trade the leading term operator $\mathcal{O}_0$ 
in Eq.~(\ref{eq:HQE-1}) by the local QCD operator $\bar Q \slashed v Q$, since its forward hadronic matrix element is normalized to unity. Expanding $\bar Q \slashed v Q$ up to the desired order in $1/m_Q$ we get
\begin{equation}
	\bar Q \slashed v Q = \mathcal{O}_0 + \tilde{C}_v \frac{\mathcal{O}_v}{m_Q} + \tilde C_\pi \frac{\mathcal{O}_\pi}{2m_Q^2} + \tilde C_G \frac{\mathcal{O}_G}{2m_Q^2} + \cdots  \,,
	\label{hqebvb}
\end{equation}
where $\tilde{C}_i$ are the matching coefficients of the full QCD current to HQET.

Finally, we use the equation of motion (EOM) of the $h_v$ field
to remove the operator $\mathcal{O}_v$ in Eq.~(\ref{eq:HQE-1})
\begin{eqnarray}
	\mathcal{O}_v =
	- \frac{1}{2m_Q} (\mathcal{O}_\pi+ c_F(\mu)\mathcal{O}_G) + \ldots\,,
	\label{LHQET}
\end{eqnarray}
where $c_F(\mu)$ is the chromomagnetic operator coefficient of the HQET Lagrangian
\begin{eqnarray}
	c_F(\mu) &=& 1 + \frac{\alpha_s(\mu)}{2\pi}\bigg[ \frac{N_c^2-1}{2N_c} + N_c\bigg(1 + \ln\bigg(\frac{\mu}{m_Q}\bigg)\bigg) \bigg]\,.
\end{eqnarray}

In order to obtain the total rate, we have to take the forward matrix element of Eq.~(\ref{eq:HQE-1}).
For this we use the full QCD states $| H_Q(p_{H_Q}) \rangle$, where $H_Q$ is the $0^-$ ground state meson
with a single heavy quark $Q$. This introduces a dependence of
the HQE parameters on the quark mass $m_Q$ through the states which is nevertheless irrelevant to the order we are working on.
The HQE parameters are defined as~\cite{Mannel:2018mqv}
\begin{eqnarray}
	\langle H_Q(p_{H_Q})\lvert \bar Q \slashed v Q \lvert H_Q(p_{H_Q})\rangle &=& 2M_{H_Q}\,,  \\
	- \langle H_Q(p_{H_Q})\lvert \mathcal{O}_\pi \lvert H_Q(p_{H_Q})\rangle &=& 2M_{H_Q} \mu_\pi^2\,, \\
	c_F(\mu)\langle H_Q(p_{H_Q})\lvert \mathcal{O}_G \lvert H_Q(p_{H_Q})\rangle
	&=& 2M_{H_Q} \mu_G^2\,,
\end{eqnarray}
where we have included $c_F(\mu)$ in the definition of the matrix element $\mu_G^2$
in order to make the HQE parameters independent of the renormalization scale $\mu$. Note that one may relate $\mu_G^2$ to the mass splitting between the ground state mesons
$H$ and $H^*$ 
\begin{equation}  
\mu_G^2 = \frac{3}{4} \Delta M_H^2 = \frac{3}{4} (M_{H^*}^2 - M_H^2)\,.
\end{equation}

\section{Outline of the calculation}
\label{sec:outline}

The first step is to insert the effective Lagrangian Eq.~(\ref{LeffEWpm}) into the optical theorem
Eq.~(\ref{eq:trans_operator}) to perform the operator product expansion and obtain the total rate in the form of
Eq.~(\ref{hqewidth}). In terms of the coefficients obtained from the matching calculation
Eqs.~(\ref{eq:HQE-1}) and (\ref{hqebvb}), in combination with the EOM Eq.~(\ref{LHQET}), we get
\begin{eqnarray}
	\label{hqedifwidth}
	\Gamma(H_Q \to X)
	&=& \Gamma^0 |V_{q_2 q_3}|^2 |V_{q_1 Q}|^2
	\bigg[ C_0 \bigg( 1
	- \frac{\bar{C}_\pi - \bar{C}_v }{C_0}\frac{\mu_\pi^2}{2m_Q^2}\bigg)
	+ \bigg(\frac{\bar{C}_G}{c_F(\mu)} -  \bar{C}_v \bigg)\frac{\mu_G^2}{2m_Q^2} + \cdots \bigg] \, ,
\end{eqnarray}
where we have defined $\bar{C}_i\equiv C_i - C_0 \tilde{C}_i$ as the difference  between the coefficients of the HQE
of the transition operator and the current multiplied by $C_0$.

The computation of the coefficients follows our previous work~\cite{Mannel:2021zzr,Moreno:2022goo} where
we take the corresponding Feynman amplitude, expand to the necessary order in the small momentum $k$, and project to
the corresponding HQET operators Eqs.~(\ref{O0}), (\ref{Ov}) and (\ref{OG}).

The Feynman diagrams contributing to the leading power coefficient $C_0$ at LO-QCD and NLO-QCD are two-loop and three-loop quark to quark self-energy-like diagrams. The ones contributing to the coefficients of power corrections $\bar C_v$ and $\bar C_G$ at LO-QCD and NLO-QCD are two-loop and three-loop quark to quark-gluon scattering diagrams.

\begin{figure}[!htb]
	\centering
	\includegraphics[width=1.0\textwidth]{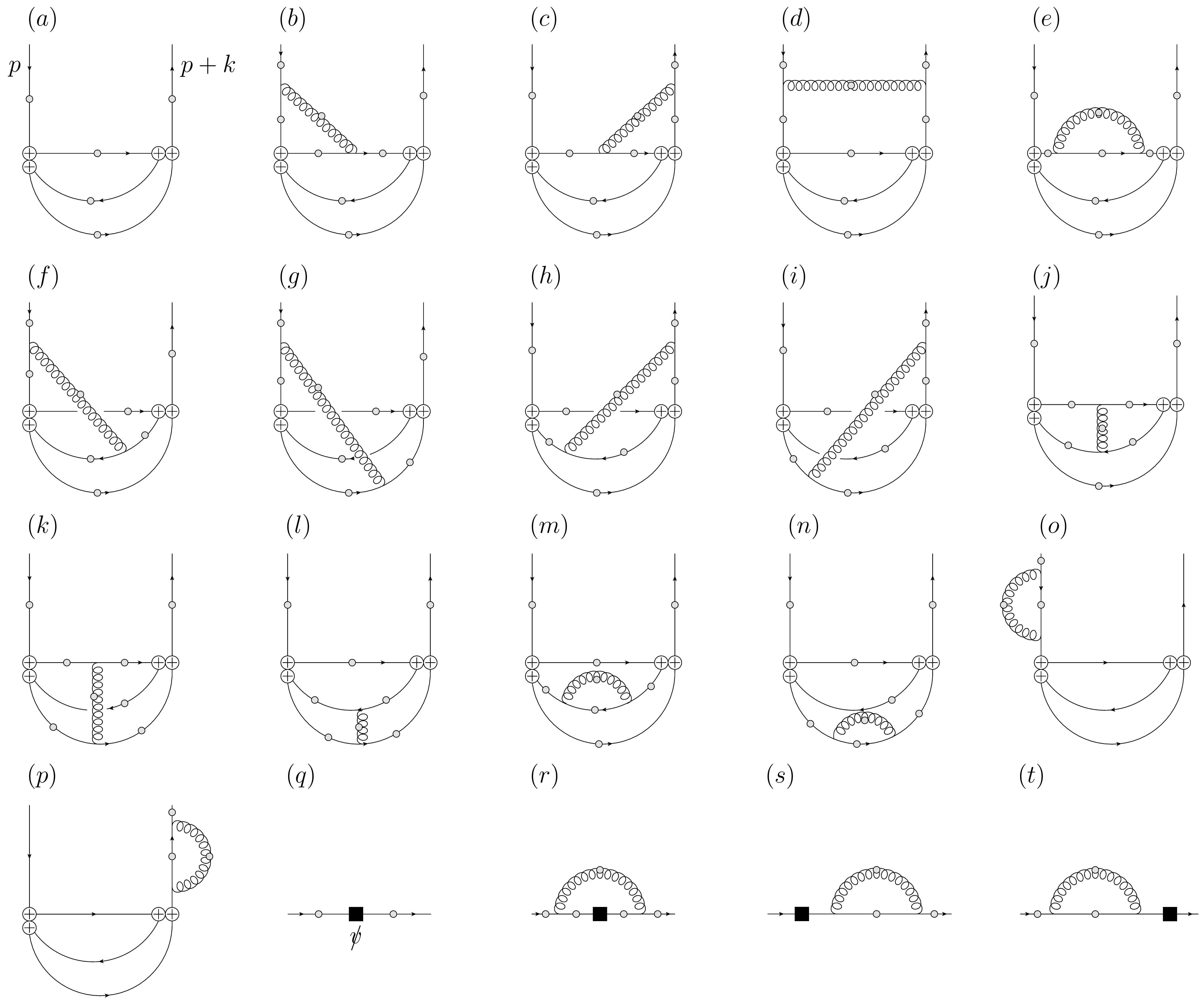}
      \caption{Feynman (a)-(t) diagrams contributing to the coefficients $C_0$, $\bar C_v$ and $\bar C_G$ of the HQE of the nonleptonic decay rate up to NLO. The incoming heavy quark carries momentum $p$, with $p^2=m_Q^2$. Grey dots stands for possible gluon insertions with incoming momentum $k\sim \Lambda_{\rm QCD}$. The black box vertex stands for $\slashed v$ insertions.
        All diagrams contributing to power corrections are obtained after taking into account all possible one-gluon insertions.
        Four-fermion vertices correspond to $\mathcal{O}_\pm$ insertions of $\mathcal{L}_{\rm eff}$.}
                \label{NLPdiagrams}
\end{figure}

The Feynman diagrams contributing to the coefficients $C_0$, $\bar C_v$ and $\bar C_G$ of the HQE of the nonleptonic decay rate up to NLO are shown in Fig.~[\ref{NLPdiagrams}]. For the computation of the leading power coefficient $C_0$ only diagrams (a-p) without gluon insertions have to be considered. For the computation of the next-to-leading power coefficient $\bar C_G$ and the auxiliary
coefficient $\bar C_v$ all diagrams (a-t) containing one-gluon insertions have to be considered.
Overall there are 14 diagrams contributing to $C_0$ up to NLO, 1 to LO and 13 to NLO.
There are 128 diagrams contributing to $\bar C_v$ and $\bar C_{G}$ up to NLO,
7 to LO and 121 to NLO.

By using LiteRed~\cite{Lee:2012cn,Lee:2013mka} the corresponding
amplitudes are reduced to a combination of the master integrals given in appendix~\ref{app:mas}.
The LO diagram Fig.~\ref{NLPdiagrams}-(a) can be reduced to the two-loop master integral Fig.~\ref{masters}-(a).
The diagrams Fig.~\ref{NLPdiagrams}-(a,e,j-n) can be reduced to a combination of the massless three-loop master integrals
Fig.~\ref{masters}-(b,c). Finally, the diagrams Fig.~\ref{NLPdiagrams}-(b-d,f-i) can be reduced to a combination
of the massive three-loop master integrals Fig.~\ref{masters}-(d,e).

We use standard dimensional regularization in $D=4-2\epsilon$ space-time dimensions with $\gamma_5$ treated in
NDR. This forces us to chose a renormalization scheme with evanescent operators preserving Fierz
symmetry to the necessary order. In this way, we can use Fierz symmetry to write all Feynman diagrams as a
single open fermionic line without $\gamma_5$ problem. Nevertheless, the explicit expressions for the coefficient functions of the HQE in terms of $C_\pm$ are scheme-dependent. This scheme dependence cancels with the corresponding scheme dependence of the coefficients
$C_\pm$.

For the algebraic manipulations including Lorentz and Dirac algebra we use Tracer~\cite{Jamin:1991dp}.
For the color algebra we use ColorMath~\cite{Sjodahl:2012nk}.
Expansion of Hypergeometric functions is done with the
help of HypExp~\cite{Huber:2005yg,Huber:2007dx}.
The computation is performed in Feynman gauge and we use the background
field method to compute the scattering in the external gluonic field.

For renormalization we adopt the $\overline{\mbox{MS}}$
renormalization scheme for the strong coupling constant $\alpha_s(\mu)$ and
the renormalization of the HQET Lagrangian.
The heavy quark is renormalized on-shell

\begin{eqnarray}
 Q_B = (Z_2^{\mbox{\scriptsize ON}})^{1/2} Q\,,\quad\quad
 Z_2^{\mbox{\scriptsize ON}}  &=& 1 - \frac{N_c^2-1}{2N_c} \frac{\alpha_s}{4\pi}\bigg( \frac{3}{\epsilon} + 6 \ln\bigg(\frac{\mu}{m_Q}\bigg) + 4 \bigg)\,.
\end{eqnarray}
Therefore, we will quote our results in the on-shell (pole mass) scheme for the heavy quark mass
$m_Q$. For most precise predictions one usually chooses for the bottom quark a low-scale short
distance mass such as the kinetic or the $1S$ mass, and thus one needs to convert
the on-shell mass into such a mass scheme for which the known one-loop expression will be sufficient.

\section{Results and discussion}
\label{sec:res}
In this section we provide the results for the coefficients of the HQE of the nonleptonic decay rate
in Eq.~(\ref{hqewidth}) up to NLO-QCD. Note that the reparametrization invariance of the HQE ensures that to all orders
in $\alpha_s (\mu)$ we have $C_0 = C_{\mu_\pi}$, so Eq.~(\ref{hqewidth}) takes the form
\begin{equation}
	\label{hqewidth-1}
	\Gamma(H_Q\rightarrow X) = \Gamma^0 |V_{q_2 q_3}|^2 |V_{q_1 Q}|^2
	\bigg[   C_0 \bigg(1
	-  \frac{\mu_\pi^2}{2m_Q^2} \bigg)
	+ C_{\mu_G}\frac{\mu_G^2}{2m_Q^2} + \cdots
	\bigg]\,,
\end{equation}
with 
\begin{equation}
	C_{\mu_G} = \frac{\bar{C}_G}{c_F(\mu)} - \bar{C}_v\,.
\end{equation} 

We show our results for the coefficients defined in Eq.~(\ref{hqewidth-1}) in the form
\begin{equation}
	C_i = C_i^{\rm LO} + \frac{\alpha_s(\mu)}{\pi}C_i^{\rm NLO}\,, \quad i = 0,\,\mu_G\,.
\end{equation}
The leading power coefficient reads
\begin{eqnarray}
	C_0^{\rm LO} &=& \frac{1}{2} N_c (C_{+}^2 + C_{-}^2)
	+ \frac{1}{2} (C_{+}^2 - C_{-}^2)
	\nonumber
	\\
	&=& \frac{3}{2}(C_{+}^2 + C_{-}^2) + \frac{1}{2} (C_{+}^2 - C_{-}^2)\,,
	\\
	C_0^{\rm NLO} &=& - (N_c^2-1) \bigg( \frac{\pi^2}{8} - \frac{31}{32} \bigg) (C_{+}^2 + C_{-}^2)
	- \frac{N_c^2-1}{2N_c} \bigg( \frac{3}{2}\ln \bigg(\frac{\mu^2}{m_Q^2}\bigg) + \frac{\pi^2}{4} + \frac{51}{16} \bigg)
	(C_{+}^2 - C_{-}^2)
	\nonumber
	\\
	&=& - \bigg( \pi^2 - \frac{31}{4} \bigg) (C_{+}^2 + C_{-}^2)
	- \bigg( 2\ln\bigg(\frac{\mu^2}{m_Q^2}\bigg)
	+ \frac{\pi^2}{3} + \frac{17}{4} \bigg)  (C_{+}^2 - C_{-}^2)\,,
\end{eqnarray}
while at subleading power we obtain   
\begin{eqnarray}
	C_{\mu_G}^{\rm LO} &=& -\frac{3}{2}N_c
	(C_{+}^2 + C_{-}^2)
	- \frac{19}{2} (C_{+}^2 - C_{-}^2)
	\nonumber
	\\
	&=& -\frac{9}{2}(C_{+}^2 + C_{-}^2) - \frac{19}{2} (C_{+}^2 - C_{-}^2)\,,
	\\
	C_{\mu_G}^{\rm NLO} &=&
	\bigg( 12\ln \bigg(\frac{\mu^2}{m_Q^2}\bigg) - \bigg( \frac{5}{288} + \frac{\pi^2}{8} \bigg) N_c^2 + \frac{31\pi^2}{24}
	+ \frac{6533}{288} \bigg) (C_{+}^2 + C_{-}^2)
	\nonumber
	\\
	&&
	+ \frac{1}{N_c}
	\bigg( \frac{3}{4}(3 N_c^2-19)\ln\bigg(\frac{\mu^2}{m_Q^2}\bigg) + \frac{13}{24}\bigg( \pi^2 - \frac{91}{12} \bigg) N_c^2
	- \frac{179\pi^2}{72} - \frac{3361}{288} \bigg) (C_{+}^2 - C_{-}^2)
	\nonumber
	\\
	&=& \bigg( 12\ln\bigg(\frac{\mu^2}{m_Q^2}\bigg) + \frac{\pi^2}{6} + \frac{811}{36} \bigg) (C_{+}^2 + C_{-}^2)
	+ \bigg( 2 \ln \bigg(\frac{\mu^2}{m_Q^2}\bigg) + \frac{43\pi^2}{54} - \frac{1751}{108} \bigg)
	(C_{+}^2 - C_{-}^2)\,,
	\nonumber
	\\
	&&
\end{eqnarray}
where in the second equalities we have replaced $N_c=3$.

Note that the coefficient functions multiplying the $C_\pm$ coefficients are in general dependent on the scheme used for
$\gamma_5$ and the choice of evanescent operators. This scheme dependence cancels with the scheme dependence of the
coefficients $C_\pm$. Therefore, the results written above together with the definitions given in Eqs.~(\ref{CpmNLO}), 
(\ref{RpmBpm}) and (\ref{lpm}) are scheme-independent. In addition we note 
that only two structures $(C_{+}^2 + C_{-}^2)$ and $(C_{+}^2 - C_{-}^2)$ appear. In
the basis of Eq.~(\ref{eq:FermiLagr}) this is translated into the two structures $(C_1^2+C_2^2)$ and $C_1 C_2$.
This is implied by Fierz symmetry.

The result obtained for the $C_0$ coefficient agrees with \cite{Bagan:1994zd} which also was obtained in NDR and using 
Fierz symmetry. This result also agrees with \cite{Altarelli:1980fi} and \cite{Buchalla:1992gc}, where this 
coefficient has been computed in dimensional reduction and the 't Hooft-Veltman scheme, respectively.

For the power suppressed terms, we re-calculated the expression obtained for the $C_{\mu_G}$ coefficient, and our result 
agrees with the result known from~\cite{Bigi:1992su,Blok:1992hw,Blok:1992he}. The new result of this calculation is the 
next-to-leading order contribution to the $C_{\mu_G}$ coefficient. 

We may also switch to a reparametrization invariant basis as discussed in      
\cite{Mannel:2018mqv}, where the HQE parameters are defined using the operators of full QCD as in Eq.~(\ref{eq:heavy-quark-me-phase})
\begin{equation} 
\langle H_Q(p_{H_Q}) | \bar{Q}_v Q_v | H_Q(p_{H_Q}) \rangle = 2 M_{H_Q} \mu_3 =
2 M_{H_Q} \bigg(1 - \frac{\mu_\pi^2 - \mu_G^2 }{2 m_Q^2} \bigg)\,.
\end{equation} 
To the order we are working on we can identify the static field with the full QCD field, and find
\begin{equation}
	\Gamma(H_Q\rightarrow X) = \Gamma^0 |V_{q_2 q_3}|^2 |V_{q_1 Q}|^2
	\bigg[   C_0 \mu_3 
	+ (C_{\mu_G}-C_0) \frac{\mu_G^2}{2m_Q^2} + \cdots
	\bigg]\,.
\end{equation} 

\begin{figure}[ht]
\centering
\subfigure[$C_0$ for bottom quark case.]{\includegraphics[scale=0.6]{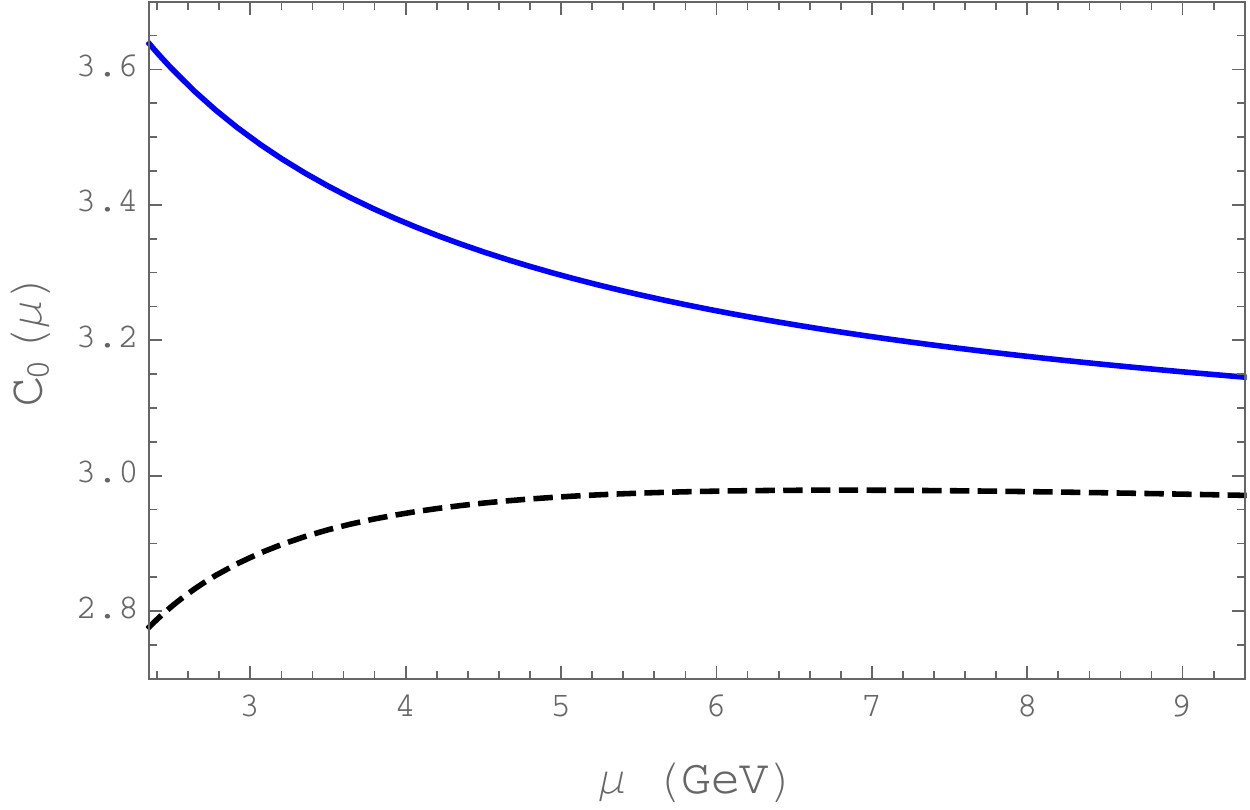}
\label{fig:C0Bottom}}
\quad
\subfigure[$C_{\mu_G}$ for bottom quark case.]{%
\includegraphics[scale=0.6]{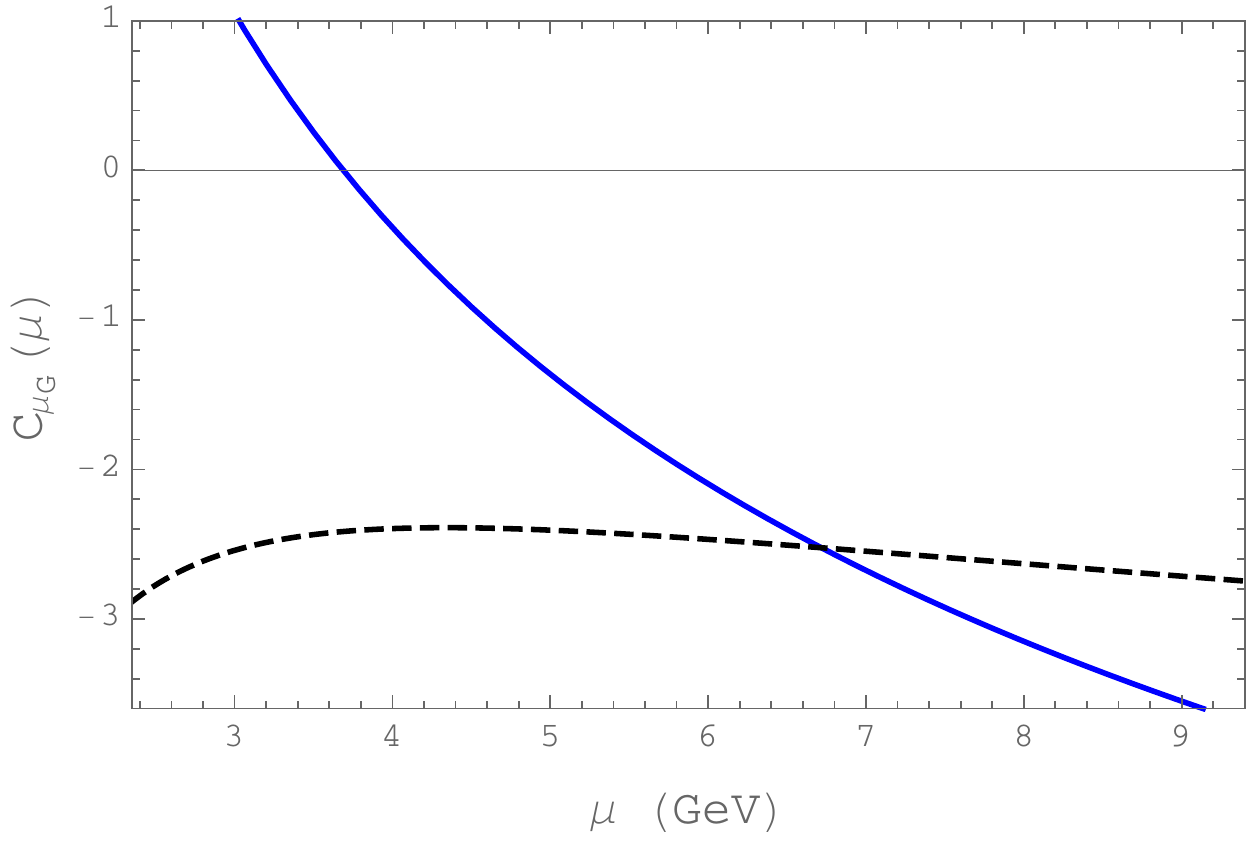}
\label{fig:CGBottom.pdf}}
\subfigure[$C_0$ for charm quark case.]{%
\includegraphics[scale=0.6]{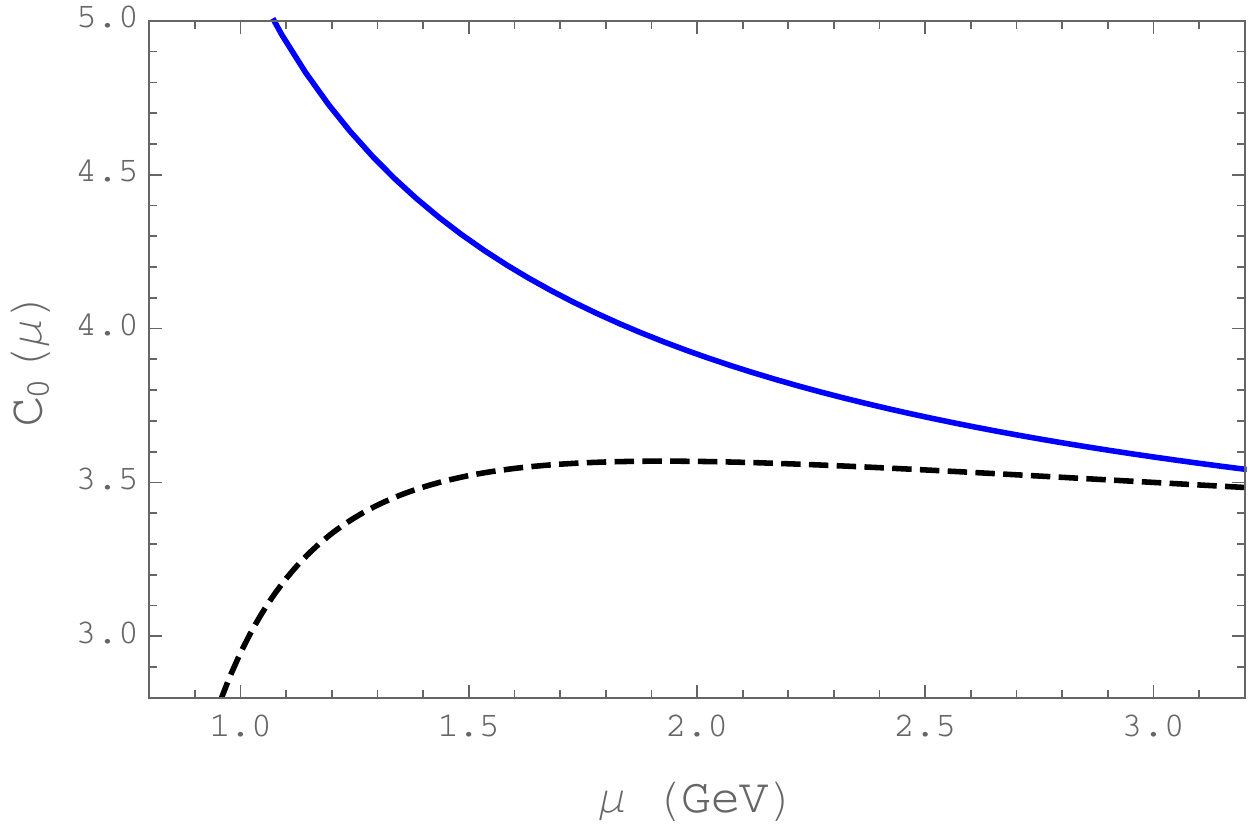}
\label{fig:C0Charm}}
\quad
\subfigure[$C_{\mu_G}$ for charm quark case.]{%
\includegraphics[scale=0.6]{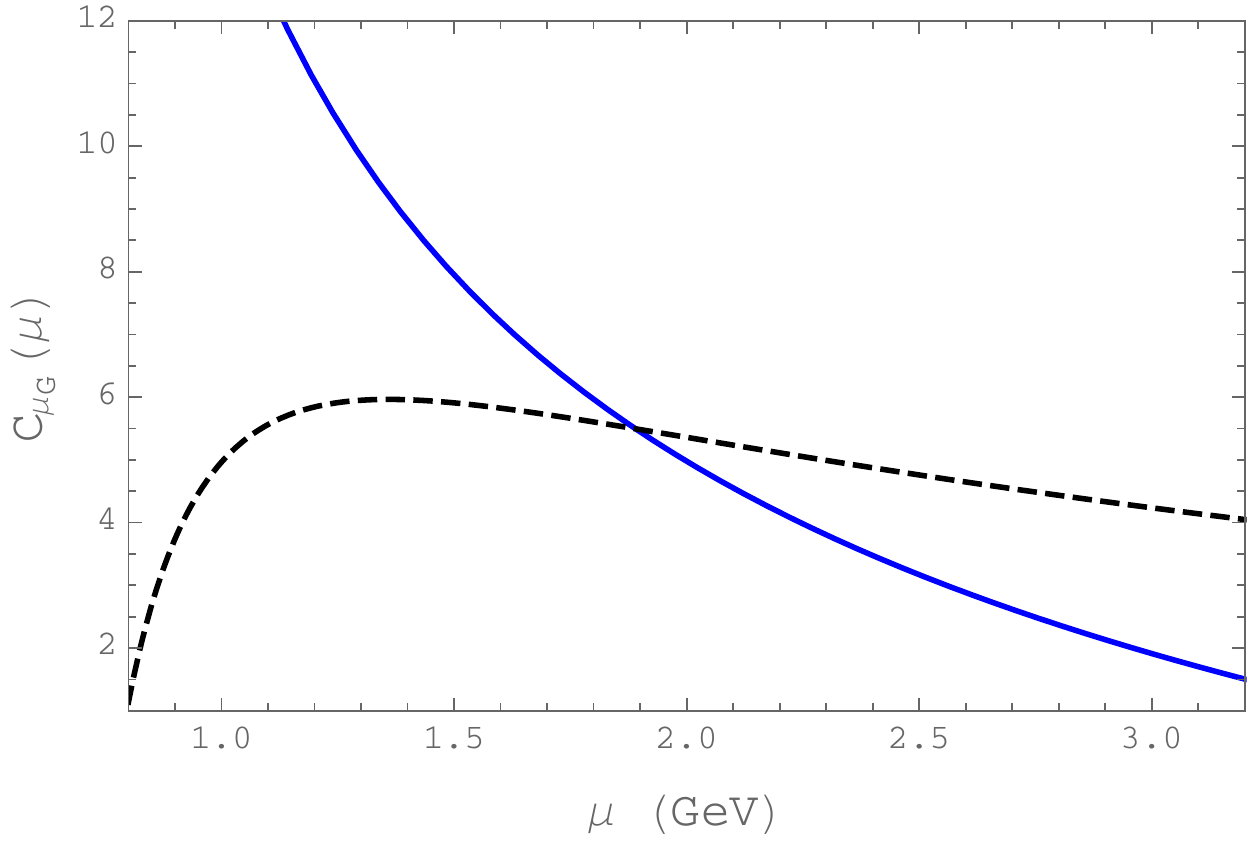}
\label{fig:subfigure4}}
\caption{The plots (a)-(d) show the dependence of the coefficients of the HQE of the inclusive nonleptonic decay rate on the renormalization scale
$\mu$ in the range $m_Q/2 < \mu < 2m_Q$. The blue continuous lines stand for coefficients with LO precision whereas the black dashed
lines stand for coefficients with NLO precision.}
\label{fig:coeffs}
\end{figure}

The NLO contributions to the coefficients are expected to reduce the dependence of the coefficients on the 
renormalization scale $\mu$, so we look at the $\mu$ dependence of $C_0$ and $C_{\mu_G}$. 
In Fig.[\ref{fig:coeffs}] we show this dependence, varying  
$\mu$ in the range $m_Q/2 < \mu < 2m_Q$ for both, the bottom- and the charm-quark cases. For illustration we take $m_b=4.7$ GeV, $m_c=1.6$ GeV, $M_Z = 91.18$ GeV and $\alpha_s(M_Z)=0.118$, from which we obtain $\alpha_s(M_W)=0.120$ at $M_W=80.4$ GeV.
For the running of the strong coupling $\alpha_s(\mu)$ we use RunDec~\cite{Chetyrkin:2000yt} to run it down from $M_W$ to $m_b$
with $n_f=5$, and from $m_b$ to $m_c$ with $n_f=4$. The two-loop running coupling is used.

As one would expect, the coefficients at NLO show a much weaker $\mu$-dependence than their LO
counterparts. This is important phenomenologically since it will allow to reduce the uncertainty due to the choice of the
scale $\mu$. This is specially true for the $C_{\mu_G}$ coefficient, where the uncertainty due to the choice
of $\mu$ is very large.

As a consequence of the strong $\mu$-dependence of $C_{\mu_G}$, NLO corrections to the $C_{\mu_G}$ coefficient are expected to 
be very large in general and should also strongly dependent on the value of $\mu$. The sum of LO and NLO contributions is, 
however, almost independent of $\mu$. Therefore, NLO corrections happen to be very important and they stabilize the numerical 
value of the coefficient.

Note that for the bottom case, the leading-order chromomagnetic operator coefficient has a zero for a value of 
$\mu \approx 3.8$ GeV, leading to a large uncertainty for this particular contribution. However, including the 
NLO contribution improves the situation significantly, leaving us with a negative contribution, lowering the
total value of the width (increase the size of the lifetimes).     

\begin{figure}[ht]
\centering
\subfigure[Bottom quark case.]{\includegraphics[scale=0.6]{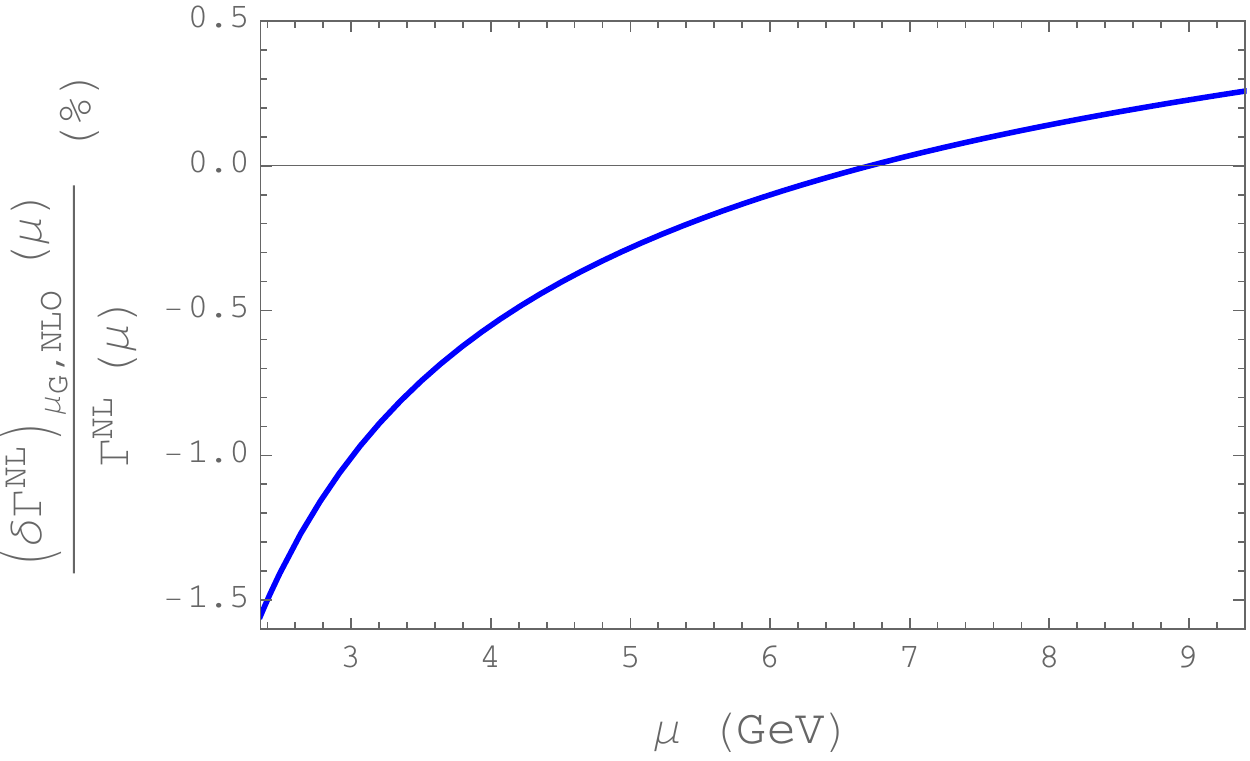}
\label{fig:dGbottom}}
\quad
\subfigure[Charm quark case.]{%
\includegraphics[scale=0.6]{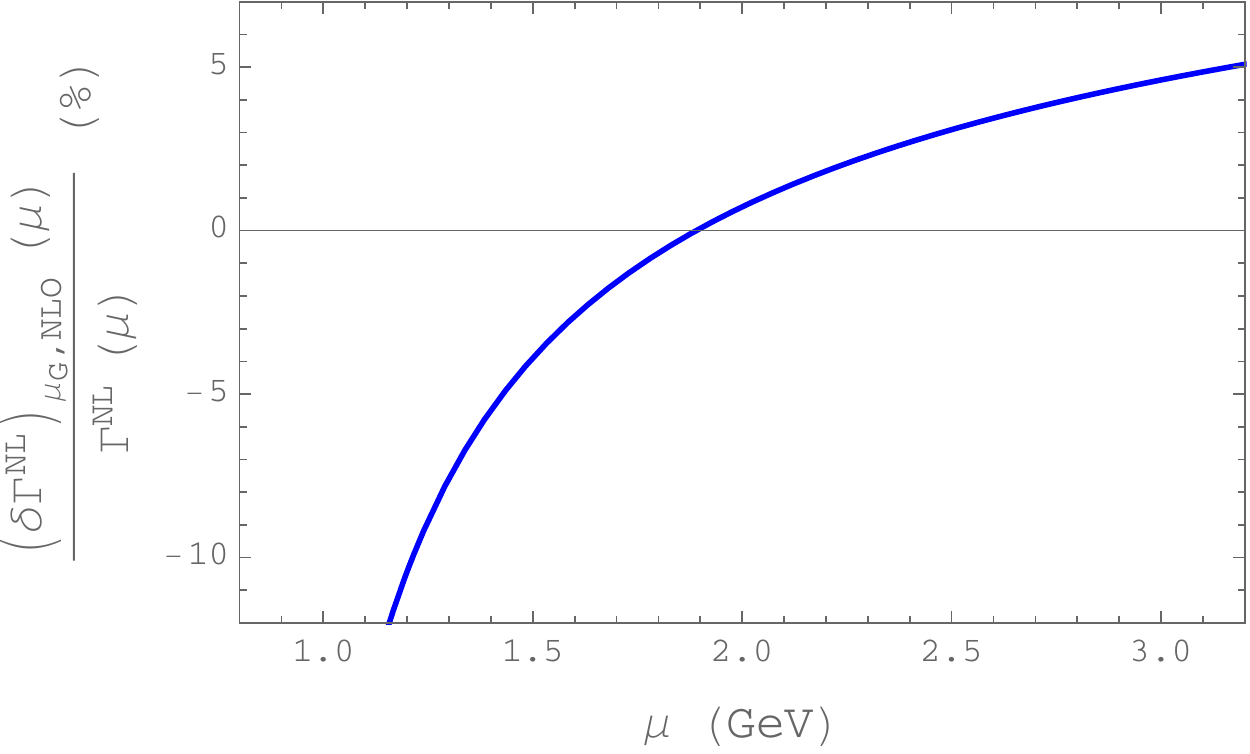}
\label{fig:dGCharm}}
\caption{Relative size between the new contribution to the nonleptonic width due to the NLO correction to the chromomagnetic operator coefficient $\delta\Gamma^{\rm NL}_{\mu_G,{\rm NLO}}(\mu)$ and the nonleptonic width $\Gamma^{\rm NL}(\mu)$ as a function of the
renormalization scale $\mu$ in the range $m_Q/2 < \mu < 2m_Q$. The left panel stands for the bottom quark case and the right panel
for the charm quark case.}
\label{fig:impact}
\end{figure}

Finally we illustrate the impact of the new contribution to the nonleptonic width by looking at the quantity 
\begin{equation}
 \frac{\delta\Gamma^{\rm NL}_{\mu_G,{\rm NLO}}(\mu)}{\Gamma^{\rm NL}(\mu)}\,,
\end{equation}
as a function of $\mu$ in the range $m_Q/2 < \mu < 2m_Q$.   
In Fig.[\ref{fig:impact}] we show its $\mu$ dependence, inserting  
 $\mu_G^2 = 0.35$ GeV$^2$ and $\mu_\pi^2 = 0.5$ GeV$^2$.
 
Based on this, we estimate a correction due to the new contribution to the nonleptonic width, and correspondingly 
to the lifetimes. We find a decrease of the rate of roughly $(-5 \pm 5)\%$ for the charm case, while the effect for
the bottom case seems to be much smaller, roughly $(-0.5 \pm 0.5)\%$. However, the bottom case has to be taken with
a grain of salt, since we did not take into account the ${\cal O} (m_c/m_b)$ effects.
 
\section{Conclusions}
\label{sec:conc}

In this paper we have computed $\alpha_s$ corrections to the chromomagnetic operator coefficient in the HQE of the
nonleptonic decay rate. This calculation represents the first attempt to include QCD corrections to power suppressed
terms in nonleptonic decays. We present an analytical result for the nonleptonic width
to order $\alpha_s(m_Q) (\Lambda_{\rm QCD}/m_Q)^2$ for the case of vanishing final state quark masses.

The main application of our result is for charm-hadron decays since our considerations correspond
to the CKM favoured decay channel $c\rightarrow s\bar d u$.
To some extent, our results can be applied to $B$ hadron decays. They constitute the main effect to order $\alpha_s(m_Q) (\Lambda_{\rm QCD}/m_Q)^2$ in the CKM favoured
decay channel $b\rightarrow c\bar u d$ up to corrections of $\mathcal{O}(m_c/m_b)$. The same is true
for the CKM favoured decay channel $b\rightarrow c\bar c s$ up to corrections of $\mathcal{O}(m_c/m_b)$ and
up to the effect of penguin operators, which are not considered in this paper.

Our main result is that the inclusion of the NLO terms significantly reduces the dependence on the renormalization
scale $\mu$. While at leading order one finds a strong dependence, including the NLO terms turns out to have almost
no $\mu$ dependence for the relevant range of $\mu$. This stabilizes the numerical predictions significantly. 

While our result can be directly applied to the case of the charm quark, where we
can safely neglect the light quark masses, the case of the bottom quark is more involved, since the charm-quark 
mass cannot be neglected and the coefficients will depend on $m_c / m_b$. It is known from the semileptonic 
case that the effects of the charm mass can be large. This will be subject of future investigations.  
          
Finally we point out that the methods used here can be extended to the next power in $1/m_Q$, i.e.
to a calculation of the NLO contribution to the Darwin operator coefficient, at least for the charm case where the
final state quarks can be treated as massless.

\subsection*{Acknowledgments}
We thank Alexander Lenz for fruitful discussions and his interest in this work.
This research was supported by the Deutsche Forschungsgemeinschaft
(DFG, German Research Foundation) under grant  396021762 - TRR 257
``Particle Physics Phenomenology after the Higgs Discovery''.

\appendix

\section{Master integrals}
\label{app:mas}

For completeness we give here the necessary master integrals for the computation of the coefficients of the HQE~\cite{Mannel:2021ubk}.
These master integrals are two- and three-loop $1\rightarrow 1$ topologies with on shell external momentum $p^2=m_Q^2$.

\subsection{Two-loop master integrals}

\begin{figure}[!htb]
	\centering
	\includegraphics[width=1.0\textwidth]{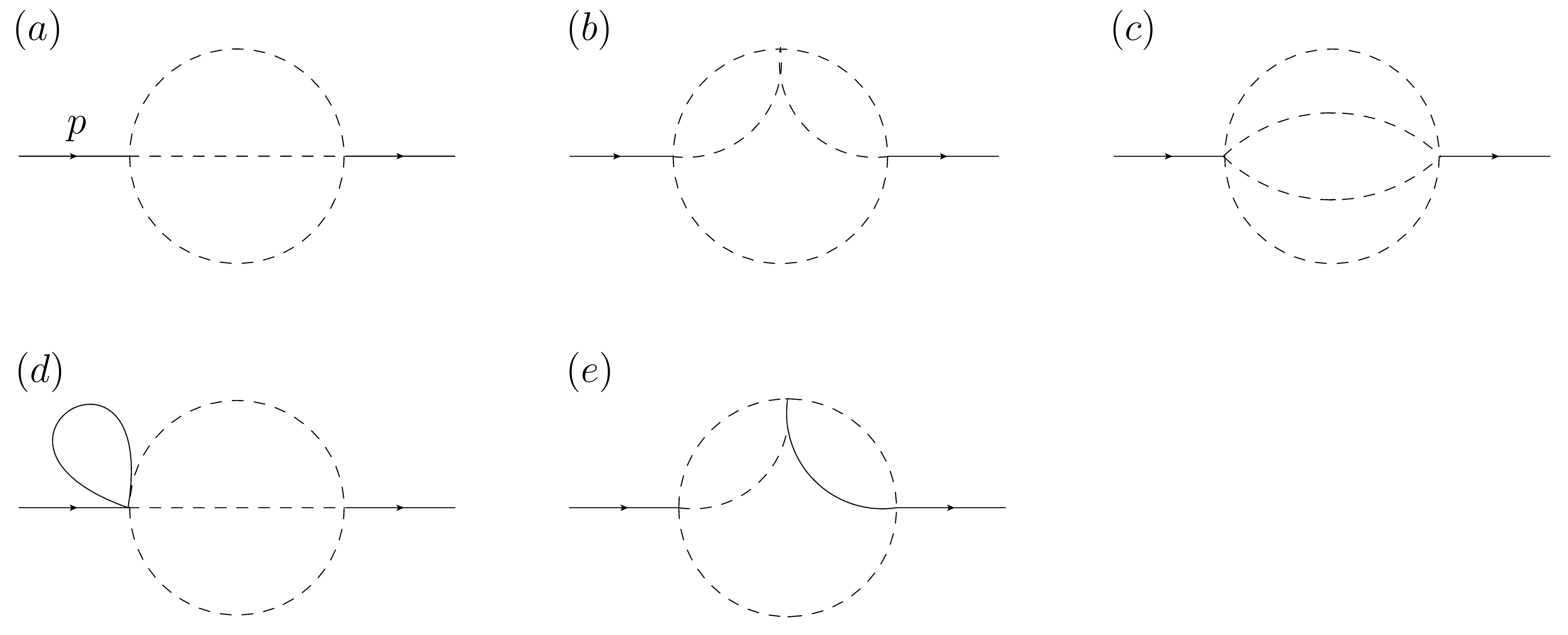}
    \caption{Master integrals (a)-(e) contributing the matching coefficients of the HQE of inclusive nonleptonic decays. The dashed lines are massless whereas the continuous lines have mass $m_Q$.}
    \label{masters}
\end{figure}

We define the following completely massless two-loop basis
\begin{eqnarray}
 && D_1 = (p-q_1)^2 \,,\quad\quad D_2 = (p-q_2)^2 \,, \quad\quad D_3 = q_1^2\,,
 \nonumber
 \\
 &&
 \quad\quad\quad\quad\quad D_4 = q_2^2\,, \quad\quad D_5 = (q_2 - q_1)^2\,.
\end{eqnarray}
To LO, the most general integral that can appear is

\begin{equation}
 \mathcal{J}(n_1,n_2,n_3,n_4,n_5) = \mbox{Im }
 m_Q^{4\epsilon} \bigg(\frac{e^{\gamma_E}}{4\pi}\bigg)^{2\epsilon}
 \int\frac{d^D q_1}{(2\pi)^D}\int \frac{d^D q_2}{(2\pi)^D}
 \prod_{i=1}^5 \frac{1}{ D_i^{n_i} }\,.
\end{equation}
After using IBP reduction only one master integral appears which is represented in Fig.~[\ref{masters}]-(a). It is a massless
two-loop sunset topology. To the necessary order in the $\epsilon$ expansion it reads

\begin{equation}
 \mathcal{J}(0,1,1,0,1) =
 \frac{m_Q^2}{512\pi^3}\bigg[
 1 + \frac{13}{2} \epsilon
 + \mathcal{O}(\epsilon^2)
 \bigg]\,.
\end{equation}

\subsection{Three-loop master integrals}

We define the following three-loop basis with one massive denominator of mass $m_Q$
\begin{eqnarray}
 && D_1 = (p-q_1)^2 \,,\quad\quad D_2 = (p-q_2)^2 \,, \quad\quad D_3 = (p-q_3)^2\,,
 \nonumber
 \\
 &&
 D_4 = q_1^2\,, \quad\quad\quad\quad\quad D_5 = q_2^2\,, \quad\quad\quad\quad\quad D_6 = q_3^2 - m_Q^2\,,
  \nonumber
 \\
 &&
 D_7 = (q_2-q_1)^2\,, \quad\quad D_8 = (q_3 - q_2)^2\,, \quad\quad D_9 = (q_3 - q_1)^2\,.
\end{eqnarray}
To NLO, the most general integral that can appear is

\begin{equation}
 \mathcal{J}(n_1,\ldots,n_9) = \mbox{Im }
 m_Q^{6\epsilon} \bigg(\frac{e^{\gamma_E}}{4\pi}\bigg)^{3\epsilon}
 \frac{1}{i}\int\frac{d^D q_1}{(2\pi)^D}\int \frac{d^D q_2}{(2\pi)^D}\int \frac{d^D q_3}{(2\pi)^D}
 \prod_{i=1}^9 \frac{1}{D_i^{n_i} }\,.
\end{equation}
After using IBP reduction four master integrals appear. Two of them are the completely massless master integrals
represented in Fig.~[\ref{masters}]-(b,c) and the other two contain one massive line of mass $m_Q$ and they are represented in
Fig.~[\ref{masters}]-(d,e). Fig.~[\ref{masters}]-(b,e) are five propagator topologies with zero and one massive lines, respectively. Fig.~[\ref{masters}]-(c) is a massless three-loop sunset topology and Fig.~[\ref{masters}]-(d) is a
two-loop sunset topology with a massive tadpole of mass $m_Q$. The explicit expressions for the master integrals to the necessary order in the $\epsilon$ expansion are
\begin{eqnarray}
 \mathcal{J}(0, 0, 1, 0, 1, 0, 1, 0, 1) &=&
 -\frac{ m_Q^4 }{49152 \pi ^5}
 \bigg[ 1
 +\frac{71}{6} \epsilon
 + \bigg( \frac{3115}{36} - \frac{7\pi^2}{4} \bigg)\epsilon ^2
 + \mathcal{O}(\epsilon^3)
 \bigg]\,,
 \\
 \mathcal{J}(0, 0, 1, 1, 1, 0, 0, 1, 1)
 &=&
 \frac{ m_Q^2}{4096 \pi ^5}
 \bigg[
  \frac{1}{\epsilon} + 10 +  \bigg( 64 - \frac{7\pi^2}{4} \bigg)\epsilon + \mathcal{O}(\epsilon^2)
   \bigg]\,,
   \\
   \mathcal{J}(0,1,0,1,0,1,1,0,0) &=&
 \frac{ m_Q^4 }{8192 \pi ^5}
 \bigg[
 \frac{1}{\epsilon }
 + \frac{15}{2}
 + \bigg( \frac{145}{4} - \frac{3\pi^2}{4} \bigg)\epsilon
 + \mathcal{O}(\epsilon^2)
   \bigg]\,,
   \\
   \mathcal{J}(0, 1, 0, 1, 0, 1, 1, 1, 0) &=&
 \frac{ m_Q^2 }{8192 \pi ^5}
 \bigg[ \frac{1}{\epsilon}
 +  \bigg(11 - \frac{\pi^2}{3} \bigg)
 + \mathcal{O}(\epsilon)
   \bigg]\,.
\end{eqnarray}

\section{The EOM operator coefficient}
\label{app:cv}

The coefficient $\bar C_v$ appears in the matching calculation of the transition operator. The corresponding operator
is redundant, and it can be removed by using the EOM. However, its coefficient is required in the
calculation as it shifts the coefficients of higher order operators. Therefore, presenting its explicit NLO expression
might be useful. We split the result as follows

\begin{eqnarray}
 \bar{C}_v = \bar{C}_v^{\rm LO} + \frac{\alpha_s(\mu)}{\pi}\bar{C}_v^{\rm NLO}\,,
\end{eqnarray}
with

\begin{eqnarray}
 \bar{C}_v^{\rm LO} &=&  \frac{5}{2} N_c (C_{+}^2 + C_{-}^2) + \frac{5}{2} (C_{+}^2 - C_{-}^2)
 \nonumber
 \\
 &=& \frac{15}{2}(C_{+}^2 + C_{-}^2) + \frac{5}{2} (C_{+}^2 - C_{-}^2)\,,
 \\
 \bar{C}_v^{\rm NLO} &=& -(N_c^2-1)\bigg( \frac{\pi^2}{8} - \frac{65}{96} \bigg) (C_{+}^2 + C_{-}^2)
 - \frac{N_c^2-1}{2N_c} \bigg( \frac{15}{2}\ln \bigg(\frac{\mu^2}{m_Q^2}\bigg) + \frac{\pi^2}{4} + \frac{1157}{48} \bigg)
 (C_{+}^2 - C_{-}^2)
 \nonumber
 \\
 &=& -\bigg( \pi^2 - \frac{65}{12} \bigg) (C_{+}^2 + C_{-}^2)
 - \bigg( 10\ln \bigg(\frac{\mu ^2}{m_Q^2}\bigg) + \frac{\pi^2}{3} + \frac{1157}{36} \bigg) (C_{+}^2 - C_{-}^2)\,.
\end{eqnarray}
Note that the color structure is the same that appears in the leading power coefficient.
The reason is that one can compute $\bar{C}_v$ by running a small momentum through the diagrams that contribute to
$C_0$, instead of considering diagrams with one-gluon insertions.


\end{document}